# Optimum design of NOLM-driven mode-locked fiber lasers

Alix MALFONDET[1], Alexandre PARRIAUX[1], Katarzyna KRUPA[2], Guy MILLOT[1,3], and Patrice TCHOFO-DINDA[1]

[1]Laboratoire Interdisciplinaire Carnot de Bourgogne, UMR 6303 CNRS, Université Bourgogne Franche-Comté, 9 Av. A. Savary, B.P. 47870, 21078 Dijon Cedex, France
[2]Institute of Physical Chemistry, Polish Academy of Sciences, Ul. Kasprzaka 44/52, 01-224 Warsaw, Poland
[3]Institut Universitaire de France (IUF), 1 Rue Descartes, Paris, France

**Most of the saturable absorbers commonly used to perform mode locking in laser cavities affect the trigger conditions of laser oscillation, which requires manually forcing the laser start-up by various means such as polarization controllers. We present a procedure for designing a laser cavity driven by a nonlinear optical loop mirror, which allows the laser to operate optimally without interfering with the oscillation triggering conditions, thus opening up possibilities for integration of this type of laser.**

The nonlinear optical loop mirror (NOLM) is well known for having enormous potential in terms of functionality and practical applications such as optical time-division multiplexing/demultiplexing [1], wavelength conversion [2], signal regeneration [3], optical modulation [4], and mode locking of lasers [5–11], to cite a few examples. Despite this impressive variety of practical applications, NOLM is still very little used in optical devices.

One of the reasons for this paradox is that this device is not always optimally configured and assembled accord- ing to the desired functionality. In fact, setting up a NOLM is a complex task, as this device affects all incident light with a multiplicity of effects whose respective consequences do not depend only on light intensity but also on the temporal profile of the input light. The magnitude of each of those effects is therefore virtually impossible to estimate in advance in many optical systems, such as mode-locked lasers and particularly in the transient regime of the laser. Moreover, most mode-locking devices, including the NOLM, have the disadvantage of interfering with the trigger conditions of the laser oscillation in the cavity. This flaw makes it necessary to use polarization controllers (PCs) to force the laser to start up [5–11]. However, recent work has unveiled a NOLM-driven mode-locked laser made entirely of polarization-maintaining (PM) fibers, which has the merit of being self-starting [12]. In the present work, we present an optimal design of mode-locked fiber lasers driven by NOLM, where one of the virtues is to preserve the auto-start feature of the laser, without using PM fibers. The architecture of a conventional NOLM (C-NOLM) consists of two major elements, namely, an asymmetric coupler and a section of loop-shaped fiber connected to the coupler [13], as schematically depicted in the dashed box in Fig. 1(a). The coupler separates the incident beam into two beams that subsequently propagate in opposite directions in the NOLM's loop, and then interfere at the output of the coupler due to the phase shift induced by self-phase modulation.

The transfer function (TF) of the C-NOLM representing the output power $P_{out}$ as a function of the input power $P_{in}$ may be expressed as :

$$P_{out} = \tau(P_{in}) \equiv \Gamma[1 - A(1 + \cos(B\,P_{in}))]P_{in} \quad (1)$$

with $A \equiv 2\rho(1-\rho)$, $B \equiv \gamma L_e(1-2\rho)/\alpha$, and $\Gamma \equiv \exp(-\alpha L)$, where $\gamma$ is the Kerr coefficient of the looped fiber, $L_e$ its effective length, $\alpha$ the attenuation coefficient including linear and connection losses, and $\rho$ is the coupler ratio. Hereafter, the fiber used as the NOLM loop is a commercial low-dispersion nonlinear fiber (LD-NLF) supplied by Sumitomo Electric Industries (SEI), having a nonlinear coeffi- cient 20 times higher than that of a standard single-mode fiber (SMF) and a second-order dispersion (SOD) coefficient ∼30 times lower than that of the SMF. To obtain more accurate TF, we performed numerical simulations, where light propagating within the NOLM's loop was modeled by a system of coupled equations of two counter-propagative waves, $E_+$ and $E_-$, given by :

$$\frac{\partial E_\pm}{\partial z} + \frac{\alpha}{2}E_\pm \pm \beta_1 \frac{\partial E_\pm}{\partial t} + i\frac{\beta_2}{2}\frac{\partial^2 E_\pm}{\partial t^2} - \frac{\beta_3}{6}\frac{\partial^3 E_\pm}{\partial t^3}$$
$$= i\gamma\left(|E_\pm|^2 + 2|E_\mp|^2\right)E_\pm, \quad (2)$$

where $\alpha, \gamma, \beta_2$, and $\beta_3$ are the coefficients of linear attenuation, nonlinearity, SOD, and third-order dispersion (TOD), respectively. $\beta_1$ denotes the inverse of the group velocity.

We have found that a NOLM can be optimally configured as a mode- locking component by suitably adjusting two key parameters, the first being the coupler ratio $\rho$, and the second being the bandwidth $\Delta\lambda_F$ of a band-pass filter (BPF) operating in synergy with the NOLM. Regarding the coupler ratio, it should be kept in mind that the NOLM divides an incident light beam into two parts, namely, a reflected beam and a transmitted beam, and in general, only one of the two beams is used in practice, while the other is suppressed by means of an optical isolator, as shown in Fig. 1(a). In fact, the value of $\rho$ determines the quantitative dis- tribution of light between the reflected and transmitted beams. The TF (1) highlights two critical situations.

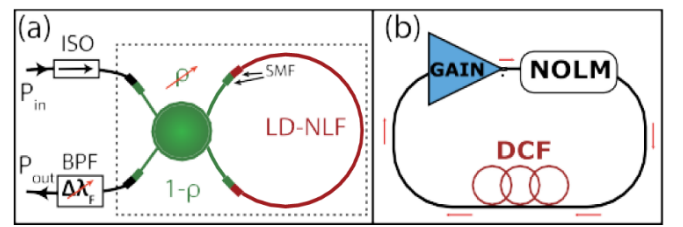

**Fig. 1.** (a) Architecture of the NOLM. $\rho$, ratio of the asymmetric coupler; BPF, band-pass filter; SMF, single-mode fiber; ISO, optical isolator; LD-NLF, low-dispersion nonlinear fiber. (b) Schematic of a NOLM-driven laser cavity.

i) If $\rho \to 1/2$, then $\tau \to 0$, and the intensity of the transmitted beam tends towards zero. The NOLM is then almost completely reflective, and the intra-cavity losses tend to infinity. The laser cannot therefore start up, regardless of the level of intra-cavity pumping.

ii) The TF (1) shows that the intensity of the transmitted beam can be increased (to the detriment of the reflected beam) by decreasing the value of $\rho$. But if $\rho \to 0$ (i.e., there is almost no reflected beam), then $\tau \to \Gamma$ and $P_{out} = \Gamma P_{in}$. The TF simply becomes a straight-line hereafter referred to as the linear transmission line (LTL). Such a TF loses a key feature of mode locking, which is to produce a saturation effect for high input powers. In this context, the following question arises : what should be the value of ratio $\rho$ to achieve mode locking without affecting the automatic start of the laser. To answer this question, it is necessary to examine more carefully the action of the NOLM at the very beginning and at the very end of the laser's transient regime. The key parameter of this approach is the slope of the TF at the origin, defined by :

$$S_0 \equiv \frac{dP_{out}}{dP_{in}}|_{P_{in} \to 0}. \qquad (3)$$

In particular, the slope of the TF (1) is given by $S_{0a} = \Gamma (2\rho - 1)^2$.

Let us start by looking at the state of the system at the very beginning of the laser's transient regime, i.e., when the laser starts up. The intra-cavity field is a photon noise, $P_{in} \ll 1$ mW and $P_{out} \ll 1$ mW, and the TF (1) becomes a straight line given by $P_{out} \sim \Gamma (2\rho - 1)^2 P_{in} = S_{0a} P_{in}$. Thus, the slope of the TF (1) at the origin, $S_{0a}$, corresponds to the power loss at each passage of the light through the NOLM at the beginning of the laser's transient regime. The power gain that fully compensates for this loss is therefore $1/S_{0a}$. Figure 2 sheds more light on the impact of the ratio $\rho$ on the general features of the TF of the C-NOLM using our LD-NLF. In particular, Fig. 2(a1) shows that in the range $0 < \rho < 1/2$, as $\rho$ increases, the slope $S_{0a}$ decreases continuously starting from a value corresponding to $\Gamma$, and tends towards zero when $\rho \to 1/2$. The curve in Fig. 2(b1) corresponds to $G_{0a} \equiv 1/S_{0a}$, i.e. the power gain needed to fully compensate for the power loss caused by the NOLM for very low input powers. As the light beam here is essentially a very weak and noisy continuous wave (CW) signal, the power loss is virtually identical to the energy loss, and the energy gain (needed to compensate for the energy loss caused by NOLM) coincides with the power gain. As mentioned above, here the TF is a straight solid line as shown in Figs. 2(c1) and 2(d1).

Let us now consider the situation where the laser is at the very end of its transient regime, and enters the stationary regime with a pulse arriving at the C-NOLM with typically: 20 W of peak power, 2 ps of temporal width, and a Gaussian temporal profile. Figures 2(a2), 2(b2), 2(c2), and 2(d2) show the main characteristics of the TF obtained from the numerical simulation of the propagation of the 20 W-2 ps pulse through the C-NOLM. In particular, Figs. 2(a2) and 2(b2), which show, respectively, the slope of the TF at the origin $S_0$ [defined by Eq. (3) and obtained from the dynamics of the parts of the pulse located in its wing ends] and the gain $G_0 \equiv 1/S_0$, coincide, respectively, with the curves of $S_{0a}$ and $G_{0a}$ in Figs. 2(a1) and 2(b1) relating to a CW of very low power. This proves that in a C-NOLM, the slope of the TF at the origin does not depend on the temporal profile of the input signal, but only on the value of ratio $\rho$. Furthermore, in Fig. 2(b2), the dashed curve represents the energy gain G that is required to compensate for the energy loss caused by the C-NOLM during the passage of the 20 W-2 ps pulse. Thus, we observe that the gain $G$ (corresponding to the laser entering the stationary regime) is well below the gain $G_0$ required for the laser to enter the transient regime.

Let us now take a closer look at what happens if tratio $\rho$ is set to 0.4, for example. In this case, a gain of 15 dB is needed to compensate for the losses caused by the NOLM at the laser start up, whereas the required gain is only 5.5 dB when the system enters the stationary regime, as shown in Fig. 2(b2). Suppose the other components in the cavity cause linear losses of a total of 6 dB (leading to 21 dB of total losses throughout the cavity when the laser starts up). If the intra-cavity amplifier can provide a gain of only 20 dB at most, then the laser will not be able to start up, and the region above 0.4 (with label "II") will be a region where the NOLM hinders laser start-up. It is precisely in this type of situation that C-NOLM prevents laser start-up, and there, PCs are often used to deform the TF until the laser starts up.

Our approach is based on the use of a variable-ratio coupler, now available on the market with a manual adjustment mechanism. By using such a coupler in the typical situation considered in Fig. 2(b2), and by gradually decreasing the value of $\rho$, we can lower the value of $G_0$ to a level that allows the laser to start up in the region labeled "I". Furthermore, Fig. 2(c2) clearly illustrates the fact that the region of small values of $\rho$ (i.e., $0 < \rho \ll 1/2$) has instead the advantage of raising the slope of the TF at the origin, which allows us to avoid any interference with the laser's auto-start functionality. However, this panel also highlights that, in the region of small $\rho$, the TF remains relatively close to the LTL, and therefore does not contain the saturation feature needed to stabilize pulses in the stationary regime. As already mentioned above, the region of large values of $\rho$ (i.e., $0 \ll \rho < 1/2$) has the disadvantage of flattening the slope of the TF at the origin, as illustrated in Fig. 2(d2); which may affect the laser's auto-start feature.

In this context, a compromise is needed between the two highlighted requirements, i.e., not to interfere with the automatic start of the laser and to produce a saturation effect at high power. In the following, we show that this compromise can be easily achieved by using the C-NOLM in synergy with a BPF, preferably a filter with an adjustable bandwidth allowing great flexibility in the setting of $\rho$.

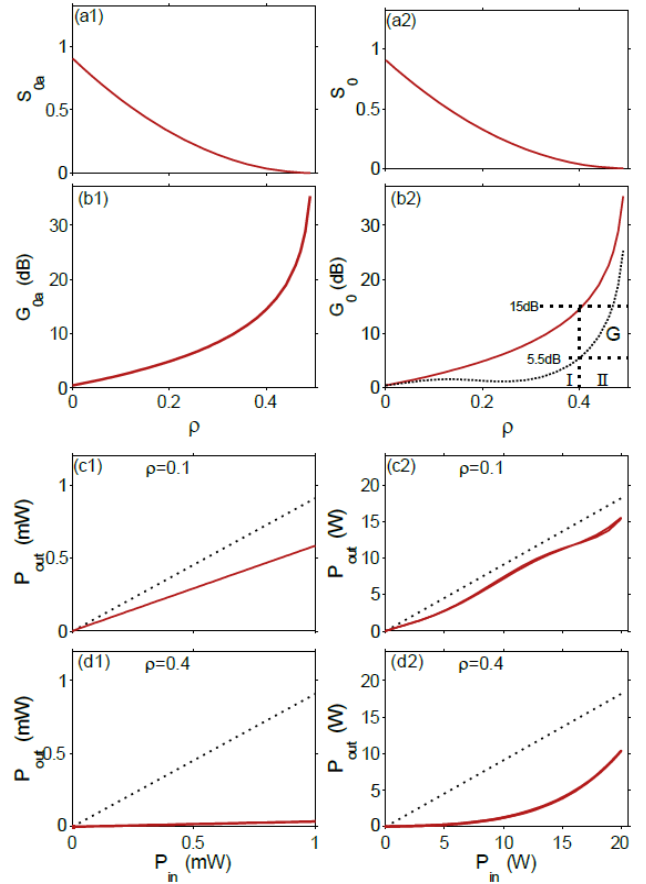

**Fig. 2.** (a1), (a2) Slope of the TF at the origin. (b1), (b2) Energy gain. (c1), (c2) TF for $\rho = 0.1$. (d1), (d2) TF for $\rho = 0.4$. The left column panels show the results of analytical calculations based on the TF (1), while the right column corresponds to numerical simulations. In (c1), (c2), (d1), and (d2), the dotted line shows LTL $P_{out} = r P_{in}$. $r = 0.9$.

The BPF can be inserted before or after the C-NOLM. However, it is preferable to insert the BPF after the exit of the C-NOLM (to maximize the power entering the NOLM). Figure 3 highlights the effects of a flat-top BPF during the passage of the 20 W-2 ps Gaussian pulse. Indeed, we found in our simulations that, qualitatively, the laser dynamics are much more dependent on the bandwidth of the BPF than on the shape of its spectral profile, and that the flat-top filter has the advantage of causing fewer losses. We modeled the flat-top BPF with a super Gaussian profile of order four. Figures 3(a1)–(a3), obtained for $\rho = 0.1$, $\rho = 0.2$, and $\rho = 0.4$, respectively, show that a 4 nm bandwidth BPF has little effect on the TF of this pulse, given the TFs in Figs. 2(c2) and 2(d2) that we obtained under the same conditions but without BPF. In contrast, Figs. 3(b1)–(b3), obtained for $\Delta\lambda_F = 2$ nm, clearly show two major effects of BPF:

(i) raising of the slope of the TF at the origin;
(ii) downward bending of the TF, which produces a saturation effect at high power, over the entire range of the values of $\rho$ considered. Thus, the filtering effect gives access to a wide window of $\rho$ values potentially allowing optimal mode locking without any interference with the laser's auto-start feature. This window extends from the region of small values of $\rho$ [Fig. 3(b1)] up to $\rho \sim 1/4$, and even beyond. However, it should be noted that choosing $\rho$ too close to 1/2 would sacrifice the pump power budget, as the losses would increase considerably due to very high reflection of the NOLM and very high absorption by the BPF, as illustrated in Fig. 3(b3).

On the other hand, band-pass filtering generates fixed points, which are surrounded by small circles in Figs. 3(b1) and 3(b2).

Here we define the fixed point as a power that, when passing through the NOLM, undergoes changes whose overall result is exactly the same as that induced by the linear losses. The presence of a fixed point offers two valuable performance characteristics, namely, a very fast transient regime when the laser starts up, and very high stability in the stationary regime.

Figures 3(c1), 3(c2) and 3(d1), 3(d2) illustrate numerical simulation of the pulse generation in our laser cavity, with the parameter sets in Figs. 3(b1) and 3(b2) for which the NOLM TF has a fixed point. Propagation in active and passive fibers (except NOLM) is described in Refs. [9,10]. We used the following parameters for the intracavity fibers, expressed (at 1.55 μm) as $P \equiv (\alpha \, [\text{dB/km}], \beta_2[\text{ps}^2/\text{m}], \beta_3[\text{ps}^3/\text{m}], \gamma \, [\text{W}^{-1} \cdot \text{m}^{-1}], L[\text{m}])$:

$P_{\text{LD-NLF}} \equiv (0.8, 7.6 \times 10^{-4}, 4.1 \times 10^{-5}, 2 \times 10^{-2}, 20)$,

$P_{\text{SMF}} \equiv (0.2, -2.3 \times 10^{-2}, 1.5 \times 10^{-4}, 1.3 \times 10^{-3}, 8.6)$,

$P_{\text{EDF}} \equiv (0.2, 1.9 \times 10^{-2}, -3.1 \times 10^{-5}, 2.9 \times 10^{-3}, 0.5)$,

$P_{\text{DCF}} \equiv (0.2, 1.9 \times 10^{-2}, 6.1 \times 10^{-5}, 2.3 \times 10^{-3}, 7)$,

where we used a fiber length $L_{\text{DCF}} = 7$ m so that the average SOD in the cavity was practically zero. In our simulations, we consider that the cavity contains only photon noise when the laser starts up. Figures 3(c1) and 3(d1) show that only a few hundred cavity round trips are sufficient for the system to reach its stationary state. Furthermore, those figures unveil that the system first generates a quasi-CW that gives rise to a large plateau within the transient regime. Next, the system goes through a relatively brief regime in which the intra-cavity field is restructured into a pulse, before entering a second plateau where the system rapidly converges towards highly stable states shown in Figs. 3(c2) and 3(d2). We also see in these two fugures that the value of $\rho$ has a very strong impact on the type of pulse profile generated. This crucial point should necessarily be taken into account when setting the value of $\rho$.

To experimentally illustrate our design approach for NOLM-based mode-locked fiber lasers, we built a fiber ring laser emitting at 1.55 μm, sketched in Fig. 4(a). We used a circulator at the entrance of the NOLM, which serves as an isolator and a cavity output. The generated pulses are characterized using their frequency-resolved optical gating (FROG) trace [14].

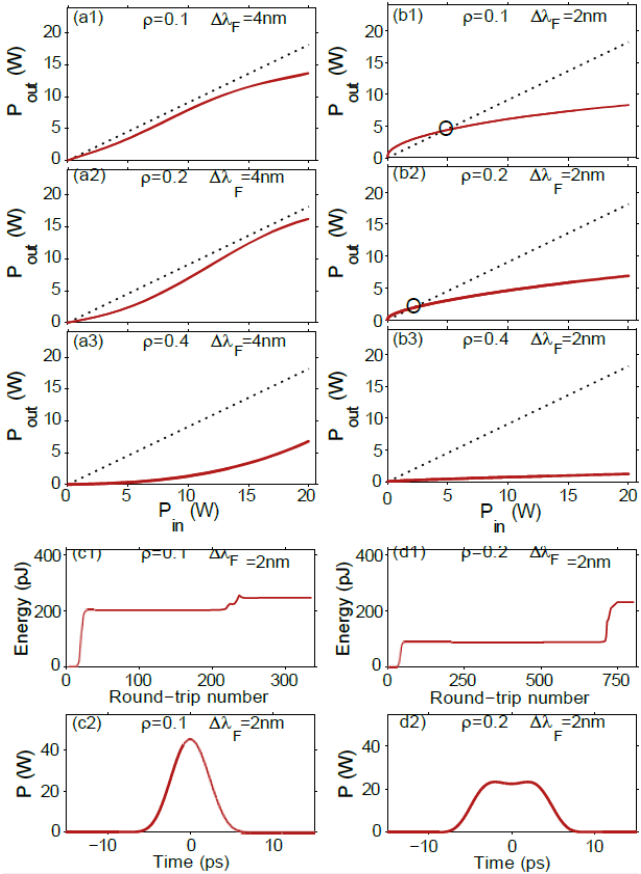

**Fig. 3.** Design of the NOLM transfer function for highly stable laser states. (a1)–(a3) and (b1)–(b3) TF for $\Delta\lambda_F = 4$ nm and 2 nm, respectively. (c1), (c2) and (d1), (d2) Pulse generation processes in the laser cavity configured using the parameter sets in (b1) and (b2), respectively.

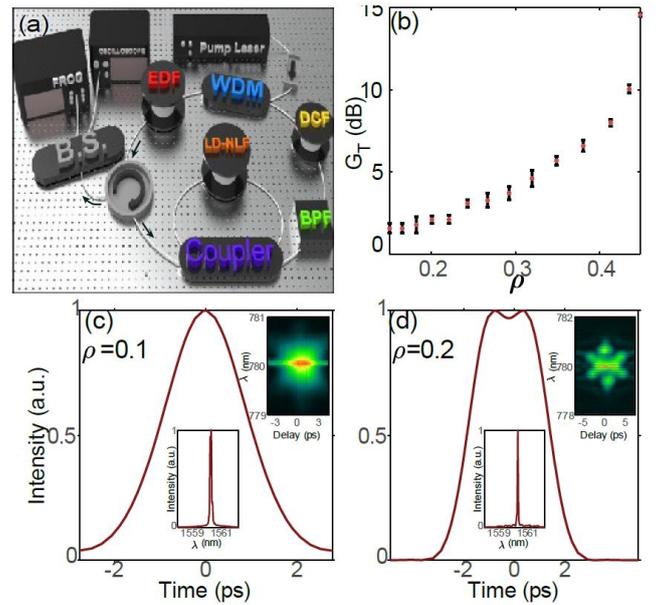

**Fig. 4.** (a) Experimental setup of the NOLM-based mode-locked fiber laser. (b) Intra-cavity gain at the mode-locking threshold. (c), (d) Intensity profile and spectrogram, respectively.

One of the key elements of our laser design approach is the NOLM's coupler. Given the crucial role of this component, it is strongly advised to check that it works properly before using it routinely. To this end, we simply proceeded in two steps. First, we calibrated our erbium-doped fiber (EDF) using a CW laser emitting at 1550 nm, whose power was set to 1 mW before injecting it into our EDF. The input and output powers of the EDF were measured using a powermeter. We thus calculated the gain G for different pumping powers $P_{pp}$ and established the calibration curve $G(P_{pp})$. Then, we installed the EDF in the cavity with all its components. We set $\rho$ to different values, and for each value, we gradually increased the pumping power (from zero) until reaching the threshold value $P_T$ triggering the mode locking. Power $P_T$ was converted into gain $G$ via the calibration curve. By doing this for different values of $\rho$, we obtained the result displayed in Fig. 4(b), which illustrates the evolution of the mode-lock gain as a function of $\rho$. This curve is consistent with the gain obtained by numerical simulation (for a laser in the stationary regime), represented by the dashed curve in Fig. 2(b2). On the other hand, Figs. 4(c) and 4(d) show the pulse profiles that we obtained by setting the coupler ratios to 0.1 and 0.2, respectively. These experimental pulse profiles are qualitatively identical to those obtained by the numerical simulations reported in Figs. 3(c2) and 3(d2).

To conclude, we have presented in this work a guideline for optimally designing NOLM-driven mode-locked fiber lasers, which is based on two key points: the first is to equip the NOLM with a coupler with a tunable ratio, $\rho$, which allows access to a variety of pulse profiles, and also serves to prevent the NOLM from interfering with the laser's self-starting conditions. The second point is to operate the NOLM in synergy with a BPF with a tunable bandwidth, $\Delta\lambda_F$, whose value is set to curve the TF of the NOLM and generate a saturation effect on high power signals. The setting of $\rho$ allows the laser dynamics to quickly evolve towards a stationary state of high stability. A NOLM- based laser designed under this guideline has many advantages compared to other mode-locked lasers such as nonlinear polari zation evolution (NPE)-based lasers [9,10].

Its first virtue is that it does not interfere with the trigger conditions of the laser oscillation within the cavity. The second is the great flexibility it offers to sculpt the TF for mode locking, by a simple calibrated adjustment of parameters $\rho$ and $\Delta\lambda_F$. This allows access to a wide variety of pulse profiles without having to go through tedious procedures of trial and error, which is required in NPE-based lasers.

As a final remark, it is worth highlighting a recent advance in nanofabrication technologies using e-beam lithography, which has made it possible to etch a NOLM on a 2.65 mm × 0.41 mm chip [15], thus proving that the NOLM is ready to go to the industrial stage. However, for NOLM-based devices to be effectively promoted as ready-to-use commercial products, it is also necessary to develop method- ologies for optimally configuring the NOLM's components according to the targeted applications. The methodology we have proposed in this study is consistent with this perspective.

**Funding**. Agence Nationale de la Recherche (ANR-15- IDEX-0003, ANR-17-EURE-0002); iXCore Research Foundation; Conseil régional de Bourgogne-Franche-Comté; Fundacja na rzecz Nauki Polskiej (POIR.04.04.00-00-16ED/18-00).

**Acknowledgment**. We thank FEMTO-EASY for the loan of the FROG system that we used in Fig. 4. K. K. acknowledges the Foundation of Polish Science.

**Disclosures**. The authors declare no conflicts of interest.

**REFERENCES**

1. T. Hirooka, D. Seya, K. Harako, D. Suzuki, and M. Nakazawa, Opt. Express **23**, 20858 (2015).
2. C. Kolleck, J. Lightwave Technol. **15**, 1906 (1997).
3. F. Wen, C. P. Tsekrekos, Y. Geng, X. Zhou, B. Wu, K. Qiu, S. K. Turitsyn, and S. Sygletos, Opt. Express **26**, 12698 (2018).
4. M. H. A. Wahid, M. M. Nahas, R. A. Ibbotson, and K. J. Blow, Opt. Commun. **332**, 55 (2014).
5. B. Ibarra-Escamilla, O. Pottiez, E. A. Kuzin, M. Duran-Sanchez, and J. W. Haus, Laser Phys. **19**, 368 (2009).
6. M. E. Likhachev, S. S. Aleshkina, and M. M. Bubnov, Laser Phys. Lett. **11**, 125104 (2014).
7. A. F. J. Runge, C. Aguergaray, R. Provo, M. Erkintalo, and N. G. R. Broderick, Opt. Fiber Technol. **20**, 657 (2014).
8. S. S. Aleshkina, M. M. Bubnov, A. K. Senatorov, D. S. Lipatov, and M. E. Likhachev, Laser Phys. Lett. **13**, 035104 (2016).
9. K. Krupa, K. Nithyanandan, U. Andral, P. Tchofo-Dinda, and P. Grelu, Phys. Rev. Lett. **118**, 243901 (2017).
10. Z. Wang, K. Nithyanandan, A. Coillet, P. Tchofo-Dinda, and P. Grelu, Nat. Commun. **10**, 830 (2019).
11. Y. X. Guo, X. H. Li, P. L. Guo, and H. R. Zheng, Opt. Express **26**, 9893 (2018).
12. J. Szczepanek, T. M. Kardas, M. Michalska, C. Radzewicz, and Y. Stepanenko, Opt. Lett. **40**, 3500 (2015).
13. N. J. Doran and D. Wood, Opt. Lett. **13**, 56 (1988).
14. D. J. Kane and R. Trebino, IEEE J. Quantum Electron. **29**, 571 (1993).
15. Z. Wang, I. Glesk, and L. R. Chen, APL Photon. **3**, 026102 (2018).